\documentstyle[12pt]{article}
\begin{document}
\thispagestyle{empty}
\noindent
\begin{flushright}
        OHSTPY-HEP-T-97-024\\
        December 1997
\end{flushright}

\vspace{1cm}
\begin{center}
  \begin{Large}
  \begin{bf}
      Gauge-mediated SUSY Breaking with a Gluino LSP
   \\
  \end{bf}
 \end{Large}
\end{center}
  \vspace{1cm}
 
    \begin{center}
    Stuart Raby\\
      \vspace{0.3cm}
\begin{it}
Department of Physics,
The Ohio State University,
174 W. 18th Ave.,
Columbus, Ohio  43210\\
raby@mps.ohio-state.edu \\
\end{it}
  \end{center}
  \vspace{1cm}
\centerline{\bf Abstract}
\begin{quotation}
\noindent
In gauge-mediated SUSY breaking models, messengers transmit
SUSY breaking from a partially hidden sector to the standard model sector
via common standard model gauge interactions.  The minimal set of messengers 
has quantum numbers of a $5 + \bar 5$ of SU(5); identical to the quantum 
numbers of the minimal Higgs sector of an SU(5) GUT.  We show in a simple model
with messenger masses of order the GUT scale that Higgs - messenger mixing 
quite naturally leads to a low energy MSSM with gluinos as the lightest 
supersymmetric particles [LSP].  We study the phenomenological consequences of 
such a model.
\end{quotation}
\vfill\eject 

\section{Gauge-mediated SUSY Breaking}
Gauge-mediated SUSY breaking [GMSB] models \cite{gmsb,dine} solve the problem of
flavor
changing neutral currents inherent in the MSSM \cite{dg,hkr}.  Consider for the
purposes of this short paper, flavor changing processes of charged leptons.
Supersymmetric charged lepton mass terms are of the form  $$ \bar e \; {\bf m}_e
\; e $$
where $ e \; (\bar e)$ represents 3 families of left-handed (right-handed)
fermions 
and their scalar partners and ${\bf m}_e$ is a complex 3 x 3 mass matrix.  In
addition, scalars necessarily have soft SUSY breaking mass terms given by
$$ \tilde e^* \;{\bf m}_{\tilde e}^2 \;\tilde e  \;\;+ \;\;  \tilde {\bar e}^*
\;{\bf
m}_{\tilde {\bar e}}^2 \; \tilde {\bar e}$$  where
$\tilde e \;(\tilde {\bar e})$ represents the left-handed (right-handed)
sleptons and 
${\bf m}_{\tilde e}^2\; ({\bf m}_{\tilde {\bar e}}^2)$ is an hermitian 3 x 3
mass
squared matrix. One may always diagonalize the supersymmetric mass term ${\bf
m}_e$ by
a simultaneous rotation of the charged lepton and slepton fields. This rotation
however
will not, in general, diagonalize ${\bf m}_{\tilde e}^2\; , {\bf m}_{\tilde
{\bar
e}}^2$, unless they are proportional to the identity matrix.  Note, off diagonal
slepton masses lead to flavor violating processes such as  $\mu \rightarrow e
\gamma,
\; \mu \rightarrow 3 e,\; \mu
\rightarrow e$ conversion, etc.  

In GMSB, SUSY breaking occurs in an almost hidden sector
of the theory due to the expectation value $F_X$, the F component of a
superfield $X$. 
Moreover, standard model [SM] squarks, sleptons and gauginos do not couple
directly to 
$X$.  Hence they do not obtain SUSY breaking masses at tree level.  The states
which
couple directly to $X$ are the messengers of SUSY breaking.  They carry SM gauge
interactions, but otherwise do not couple to squarks and sleptons directly. 
Thus SUSY
breaking enters the SM sector at one loop to gauginos and at two loops to
squarks and
sleptons.  These SUSY breaking effects are dimensionally of order  $\Lambda
\equiv
F_X/M$ where $M$ is the messenger mass.  Moreover, they are determined by gauge
quantum
numbers; thus, for example, the  matrices ${\bf m}_{\tilde e}^2, \; {\bf
m}_{\tilde
{\bar e}}^2$ are proportional to the identity matrix at $M$.  As a result
individual lepton number is conserved.  Hence processes such as  
$\mu \rightarrow e \gamma$ are forbidden.\footnote{Our discussion ignored the
possibility of new  flavor
violating interactions due to physics at the GUT scale, $M_G$.  These
interactions can only enter through loops containing GUT mass states,  hence
they generate off diagonal mass squared terms suppressed by
factors of $(M/M_G)^2$.}

\section{The Minimal Messenger Sector}

The messenger states must carry both color and electroweak quantum numbers.  In 
addition, the messengers should be in complete SU(5) representations, to
preserve GUT
predictions for gauge couplings.  The minimal set of states satisfying
these criteria transform as a  $5 + \bar 5$  with the color triplet (weak
doublets) 
denoted as follows $ t, \; \bar t$ ($d,  \; \bar d$).  In the minimal models,
all
messengers have a common mass M.   The resulting soft breaking masses are as
follows.

Gauginos obtain mass at one loop given by
\begin{eqnarray}
    m_{\lambda_i} = &  {\alpha_i(M)\over4\pi} \Lambda & ( {\rm for}\;\; i =
1,2,3). 
\end{eqnarray}

The scalar masses squared arise at two-loops
 \begin{eqnarray} \tilde m^2 = & 2 \Lambda^2 \left[
\sum_{i=1}^3 C_i\left({\alpha_i(M) \over 4 \pi}\right)^2\right] & 
  \label{eq:mtilde} \end{eqnarray}
where $C_3 = {4 \over 3}$ for color triplets and zero for singlets, 
$C_2= {3 \over 4}$ for weak doublets and zero for singlets, and $C_1 = 
{3 \over 5}{\left(Y\over2\right)^2}$, with the ordinary hypercharge $Y$
normalized as $Q = T_3 + {1 \over 2} Y$  and $\alpha_1$,  GUT normalized.   

In the limit $M << M_{G}$ ($M_G$ is the GUT scale), 
we have $\alpha_3(M) >> \alpha_2(M) > \alpha_1(M)$. Thus right-handed sleptons
are 
expected to be the lightest SUSY partners of SM fermions and binos are the
lightest
gauginos.

\section{SUSY GUT and Higgs - Messenger Mixing}

In the minimal SU(5) SUSY GUT, Higgs doublets are contained in a 
$5_H + \bar 5_H$. In order to avoid large baryon number violating nucleon 
decay rates, the color triplet Higgs $t_H, \; \bar t_H$ must have mass of 
order $M_G$, while the Higgs doublets $d_H, \;\bar d_H$ remain massless at 
the GUT scale.   The latter are responsible for electroweak symmetry breaking 
at $M_Z$.  

Our main observation\cite{raby} is that the Higgs in a SUSY GUT and the
messengers 
of GMSB have identical quantum numbers.   Thus, for messengers with mass at an
intermediate scale, 
Higgs-Messenger mixing is natural.  Moreover  as a result of doublet-triplet
splitting
in the Higgs sector, the  doublet and triplet messengers will also be split. 
This can
have significant consequences for SUSY breaking masses.

As a simple example, consider the natural doublet-triplet splitting 
mechanism in SO(10) \cite{dw}.  
The $10$ of SO(10) decomposes into a  $5 + \bar 5$ of SU(5) and the 
adjoint $45$ can be represented by an anti-symmetric 
$10 \times 10$ matrix.  The Higgs sector superspace potential is given by
\begin{eqnarray}
W_{Higgs} &  = 10_H \; 45 \; 10 \;\; + \;\; X \; 10^2 & 
\end{eqnarray}
where $10_H$ contains $5_H + \bar 5_H$, $10$ is an auxiliary 
$5 + \bar 5$ introduced for doublet-triplet splitting and $X$ is a singlet. 
Assuming 
$< 45 > = M_G (B - L)$, i.e. $45$ obtains an  SO(10)
breaking vacuum expectation value in the $B - L$ direction and $< X > = M$ , 
we obtain the triplet (doublet) mass terms given by
\begin{eqnarray}
W_{Higgs} & = t_H \; M_G \; \bar t \;\; + \;\; t \; M_G \; \bar t_H \;\; + 
\;\; M \; t \; \bar t  & \nonumber \\
 & + \;\; M \; d \; \bar d  &
\end{eqnarray}

Note, the triplets naturally have mass of order the GUT scale, while the
auxiliary 
doublets have mass $M$ and the Higgs doublets are massless.\footnote{There are 
several different ways that a $\mu$ term for the Higgs doublets can be generated

once SUSY is broken. We will not discuss this issue further here.}  The doublet
mass
is necessarily smaller than the triplets in order to suppress baryon number
violating interactions \cite{babubarr}.  Specifically, if only $10_H$ couples
to quarks and leptons, then the effective color triplet Higgs mass $\tilde M_t$ 
which enters baryon decay amplitudes is given by  $\tilde M_t = M_G^2/M$.  Hence
$\tilde M_t > M_G$ implies $ M/M_G < 1$.

The theory we propose, with Higgs-messenger mixing, is quite simple. Assume the
auxiliary $10$ is the messenger of SUSY breaking, i.e.  assume that $X$ gets
both
a SUSY conserving vev $M$  and SUSY breaking  vev $F_X$ --
\begin{eqnarray}
< X > & =  M  \;\; + \;\;  F_X \; \theta^2 ;  &  \nonumber \\
  \Lambda \;\; =  \;\;{F_X \over M} & \sim \; 10^5 \;\; \rm GeV ; &  \nonumber
  \\
  A \;\; \equiv \;\; {M \over M_G} & \sim \; 0.1  & .
\end{eqnarray}
Since the triplet messengers (mass O($M_G$)) are heavier than the doublets
(mass O($M$)), {\em SUSY 
breaking effects mediated by color triplets are suppressed.}  This has 
significant consequences for gluinos which only receive SUSY violating 
mass corrections through colored messengers.

Gauginos obtain mass at one loop given by
\begin{eqnarray}
    m_{\lambda_i} = & D_i  {\alpha_i(M)\over4\pi} \Lambda + 
    {\alpha_i(M)\over2\pi} 
    \Lambda B^2 & \;\;\;( {\rm for}\;\; i = 1,2,3)   \label{eq:gluino}
\end{eqnarray}
where $D_1 = {3 \over 5},\; D_2 = 1,\; D_3 = 0$.  

\noindent
$\bullet$  In order to 
generate SUSY violating gaugino masses, both SUSY and R symmetry must be 
broken.  In this theory,  $F_X$ breaks SUSY and the scalar vev $M$ breaks the
R symmetry which survives GUT symmetry breaking.  Thus both are necessary 
to generate the SUSY violating effective mass operator given by
\begin{eqnarray}
 &  {1 \over {\cal M}^2} \int d^4\theta \; X^{\dagger} \, X \;
 W^{\alpha}_i\,W_{\alpha \; i}
 & \;\;{\rm for}\;\; i = 1,2,3 \end{eqnarray}
 where ${\cal M}$ is determined by the heaviest messenger entering the loop.
 
\noindent
$\bullet$ Note the terms proportional to $B^2$.  Without them the gluino mass
vanishes
 at one loop due to an accidental cancellation.\footnote{I thank
 Kazuhiro Tobe for pointing this out to me.  Note, eqns.
(\ref{eq:gluino},~\ref{eq:mtilde2}) are corrections for similar equations in
ref.
~\cite{raby}. }  In order to compensate for this
 one loop cancellation we include additional messengers with a common
 mass of order $M_G$, and an R symmetry breaking mass $M$. 
 This sector thus contributes a common mass correction proportional to $B \sim
 M/M_G$.

The scalar masses squared arise at two-loops.  We obtain
 \begin{eqnarray} \tilde m^2 = & 2 \Lambda^2 \left[
 C_3\left({\alpha_3(M) \over 4 \pi}\right)^2 (A^2 + 2\; B^2) + 
C_2\left({\alpha_2(M) \over 4 \pi}\right)^2 (1 + 2\; B^2)\right] & \nonumber \\ 
& + 2 \Lambda^2 \left[ C_1\left({\alpha_1(M) \over 4 \pi}\right)^2 ({3 \over 5}
+ 
{2 \over 5}\;A^2 + 2\; B^2) \right] &
 \label{eq:mtilde2} \end{eqnarray}
where $C_i, \;\; i \; = \; 1, 2, 3$ are defined after equation ~\ref{eq:mtilde}.

\section{Low Energy Spectrum}

The heaviest SUSY particles are electroweak doublet squarks and sleptons 
and weak triplet winos, while right-handed squarks, sleptons and binos 
are lighter.  Finally gluinos are expected to be the lightest SUSY 
particle [LSP].   We have the approximate mass relations\footnote{neglecting
terms of order
$A^2$ or $B^2$, when possible},
after renormalization group running to $M_Z$,
\begin{eqnarray}
 M_2 & = {\alpha_2(M_Z)\over4\pi}\; \Lambda  &\; \approx 3 \times 10^{-3} 
 \;\Lambda \nonumber \\ 
 M_3 & = {\alpha_3(M_Z)\over2\pi}\;  B^2 \; \Lambda  &\;
\approx 9 \times 10^{-5}
\;(B/0.1)^2\; \Lambda .
\end{eqnarray}
In addition, the gravitino mass (which sets the scale for supergravity mediated
soft SUSY
breaking effects) is given by  
\begin{eqnarray}
m_{3/2} & =  ({F_X \over \sqrt{3} M_{pl}}) . & 
\end{eqnarray}
Hence
\begin{eqnarray} 
m_{3/2} & =  ({M_G \over \sqrt{3} M_{pl}}) \; B \;\Lambda  &\;\approx 6 \times
10^{-4} \;
(B/0.1) \; \Lambda .
\end{eqnarray}
The gluino mass $M_3$ depends on the arbitrary parameter $B$, the 
ratio of the R symmetry breaking scale $M$ to the typical messenger mass 
of order $M_G$.  The gravitino mass also depends parametrically on $B$ when
expressed in terms of the SUSY breaking scale $\Lambda$.

With $\Lambda = 10^5$ GeV, we obtain
\begin{eqnarray}
 M_2 &   \approx 300\;\; \rm GeV  & \nonumber \\ 
m_{3/2} &  \approx 60 \; (B/0.1) \;\; \rm GeV & \nonumber \\ 
M_3 &  \approx 9  \;(B/0.1)^2\;\; \rm GeV  &
\end{eqnarray}

\section{ Signatures of SUSY with a Gluino LSP}

Gluinos are stable.\footnote{Assuming R parity is conserved.}
They form color singlet hadrons, with the lightest of them\cite{farrar} 
given by
\begin{eqnarray}
R_0 & = \tilde g \; g  & \nonumber \\
\tilde \rho & =  \tilde g \; q \; \bar q & \;\; {\rm with}\;\; q = u,\; 
d \nonumber \\
S_0 & = \tilde g \; u \; d\; s  & 
\end{eqnarray}
where  $R_0$ is an iso-scalar fermion [glueballino]; $\tilde \rho$ 
is an iso-vector fermion and $S_0$ is an iso-scalar boson with baryon number
1.

It is unclear which one is the stable color singlet LSP.  For this paper, 
we assume $R_0$ is lighter and that both $\tilde \rho, \;\; S_0$ 
are unstable, decaying via the processes $\tilde \rho \rightarrow R_0 + \pi$ 
and $S_0 \rightarrow  R_0 + n$.

Consider the consequences of a gluino LSP.  First, the missing energy signal 
for SUSY is seriously diluted. An energetic gluino, produced in a high energy
collision, will fragment and form an hadronic jet containing an $R_0$.  
The $R_0$ will deposit energy in the hadronic calorimeter.   Thus collider
limits on squark and gluino masses must be re-evaluated.   Gluinos with mass as 
large as 50 GeV may have escaped detection.\footnote{This is a rough estimate 
using CDF limits on squarks and gluinos based on missing energy.  
Clearly if the gluino is heavy enough it will leave a
missing energy signal.}   A lower bound  on the gluino mass of 6.3 GeV has 
been obtained using LEP data on the running of $\alpha_s$ from $m_{\tau}$ 
to $M_Z$.\cite{fodor}  Thus glueballinos
may be expected in the range from  6 -- 50 GeV.

At LEP a 4 jet signal is expected  above the squark threshold, since 
squarks decay into a quark plus gluino.

Now consider possible constraints from 
exotic heavy isotope searches.   Stringent limits exist on heavy isotopes of
hydrogen. 
However, an $R_0$ must be in a bound state with a proton in order for these
searches to be relevant. Such a bound
state is unlikely due to the short range nature of the interaction of $R_0$ with
hadrons;  predominantly due to the exchange of a glueball (the lightest of which
is 10 times
heavier than a pion) or
 multiple pions.  Strong limits on heavy isotopes of oxygen also exist.
An $R_0$ can certainly be trapped in the potential well of a heavy nucleus.
However, for these searches to be restrictive,  the expected abundance of
$R_0$-nucleus bound states must be above the experimental bounds. The dominant
process 
for forming such bound states is for
$R_0$s, produced by cosmic ray collisions in the earth's
atmosphere, to be captured into nuclei.  A back of the envelope
estimate gives an expected abundance bordering on the
observable limit.  A more detailed calculation is therefore needed to say more.

Finally, what about the cosmological abundance of
$R_0$s.   Since $R_0$s annihilate via strong processes  $R_0 \; + \; R_0 
\rightarrow  2\; \pi$, the cosmological abundance is quite
 suppressed.  A rough estimate gives  
\begin{eqnarray}
n_{R_0}  &  = 10^{-10} \;\; ({m_{R_0} \over m_{\pi}}) \;\; n_B  &  
\end{eqnarray}
where  $n_B$ is the cosmological baryon density.   As a result, $R_0$s are
NOT dark matter candidates.

{\bf Acknowledgements}
This paper greatly benefited from extensive discussions with  M. Carena, 
C.E.M. Wagner, G.G. Ross, J. Gunion, H. Haber, C. Quigg, X. Tata, H. Baer, 
U. Sarid, K. Tobe, A. Mafi, S. Katsanevas, E.W. Kolb, A. Riotto, L. Roszkowski, 
G. Steigman, R. Boyd, A. Heckler, and the hospitality of the Aspen Center for 
Physics.  Finally, this work is partially 
supported by DOE grant DOE/ER/01545-730.

%

\end{document}